\documentclass[twoside,a4paper]{article}
\usepackage{fleqn,epsf,graphicx}
\usepackage{authblk}

\begin{document}

\title{Project of the underwater system for chemical threat detection}

\author{M.~Silarski$^{a}$, D.~Hunik$^{a}$, P.~Moskal$^{a}$, M.~Smolis$^{a}$, S.~Tadeja$^{a}$}

\affil{$^{a}$Institute of Physics, Jagiellonian University \\ {\L}ojasiewicza 11, 30-348 Krak\'ow, Poland}

\maketitle                   


\begin{abstract}
In this article we describe a novel method for the detection of explosives and other
hazardous substances in the marine environment using neutron activation.
Unlike the other considered methods based on this technique we propose to use guides for
neutron and gamma quanta which speeds up and simplifies identification. Moreover, it may
provide a determination of the density distribution of a dangerous substance.
First preliminary results of Monte Carlo simulations dedicated for design of a device
exploiting this method are also presented.
\end{abstract}
\section{Introduction}
During both World Wars the Baltic Sea was a scene of intense naval warfare and became
a ``container'' for almost whole Nazi chemical munition arsenal.
In accordance with the provisions of the Potsdam Conference, Germany was demilitarized
but only a small part of their arsenal was neutralized on land. Technologies
for safe and effective disposal of chemical weapons were not known shortly after the war.
Thus, till 1948 about 250000 tons of munition, including up to 65000 tons of chemical agents,
were sunk in Baltic Sea waters by both Germans and Allies. The main known
contaminated areas are Little Belt, Bornholm Deep (east of Bornholm) and the south-western
part of the Gotland Deep~\cite{czasmorza}. Apart from known underwater stockyard there
is unknown amount of dangerous war remnants spread over the whole Baltic, especially
along maritime convoys paths and in vicinity of coasts.
It is not clear how dangerous are these underwater arsenals. At the bottom
of the sea (in approx. 5-7$^o$C) the chemical agents take a form of oily liquids hardly soluble
in water. Thus, the sunk ammunition does not release hazardous substances. It becomes
dangerous however, if the rusted tanks and shells are raised from the bottom of the sea.
Chemical munitions, containing mostly mustard gas, was fished several times by fishermen
on the Baltic Sea over the last fifty years. Moreover, already in 1952 and 1955 the contamination
was found at the polish coast causing serious injuries to people.
It was estimated that if only $1/6$ of the sunk chemical agents was released
into Baltic the life in the sea and at its shores would be entirely ruined for the next 100
years~\cite{czasmorza}. High economic and environmental costs have been preventing so far any activities aiming
at extraction of these hazardous substances, but it is clear that we are about to face a very
serious problem in the Baltic Sea. Appropriate actions for preventing the ecological catastrophe
demand a precise knowledge of location and amount of sunk munitions.\\
Presently used methods for underwater munition detection is based on sonars which show only a shape
of underwater objects, like e.g. sunk ships or depth charges. To estimate the amount of dangerous
substances and to determine the exact location of sunk munition it is still necessary that people
are diving
and searching the bottom of the sea. This operation is always very dangerous for divers
since the corrosion state of the shells is usually not known. Moreover, these methods are
very expensive and slow, thus they cannot be used in practice to search big sea areas.
The above mentioned disadvantages can be to large extend overcome by using devices based on
Neutron Activation Analysis techniques (NAA) which will be discussed in more details in next
sections of this article.
\section{Underwater detection of hazardous substances with neutron beams}
Most of the commonly used explosives or drugs are organic materials. Therefore,
they are composed mostly of oxygen, carbon, hydrogen and nitrogen. War gases contain also
sulfur, chlorine, phosphorus and fluorine.
Thus, these substances can be unambiguously identified by the determination of the ratio
between number of C, H, N, O, S, P and F atoms in a molecule, which can be done noninvasively
applying Neutron Activation Analysis techniques.
The suspected item can be irradiated with a flux of neutrons produced using compact
generators based e.g. on deuteron-tritium fusion, where deuterons are accelerated to the energy
of 0.1 MeV and hit a solid target containing tritium. As a result of the fusion an alpha particle
is created together with the neutron, which is emitted nearly isotropically with a well defined
energy equal to about 14.1 MeV~\cite{moskalAnn}.
Such energy is sufficient to excite all nuclei composing organic substances and
the resulting $\gamma$ quanta from  the de-excitation of nuclei are then detected by e.g. a germanium
detector providing a very good energy resolution. Counting the number
of gamma quanta emitted by nuclei from the examined item provides information about its stoichiometry.
Devices using NAA to detect explosives on the ground were already designed
and are produced, e.g. in USA~\cite{maglich} and Poland~\cite{SWAN}.
In the aquatic environment however we encounter serious problems since neutrons are strongly
attenuated by water. Moreover, as in the case of ground detectors, an isotropic generation
of neutrons induces a large environmental background, in this case from oxygen. This noise can be
significantly reduced by the requirement of the coincident detection of the alpha particle,
which allows for the neutron tagging~\cite{mpdActab}. The attenuation of neutrons can be
compensated by decreasing the distance between generator and examined item~\cite{uncoss}.
There are also solutions based on low energy neutrons which are moderated in water before
reaching the tested object. The detector is then counting the gamma quanta from thermal
neutron capture and secondary neutrons originating from the irradiated object.
The identification is done by searching for anomalies in the observed spectra of gamma quanta
and neutrons~\cite{pat}.
However, these methods do not allow to detect explosives buried deeply in the bottom of the sea.
Moreover, the device has to approach the suspected item very close and the strong attenuation of
neutrons and gamma quanta significantly increases the exposure time and make the interpretation of
results difficult.
Therefore, we propose to build a detector which uses NAA technique and special guides
for neutrons and emitted gamma rays~\cite{patent}. The device allows for detection of
dangerous substances hidden deep in the bottom of the sea with significantly reduced
background and provides determination of the density distribution of the dangerous
substance in the tested object.
\section{Concept of using the neutron guides}
\begin{figure}
\centering
\includegraphics[width=9.0cm]{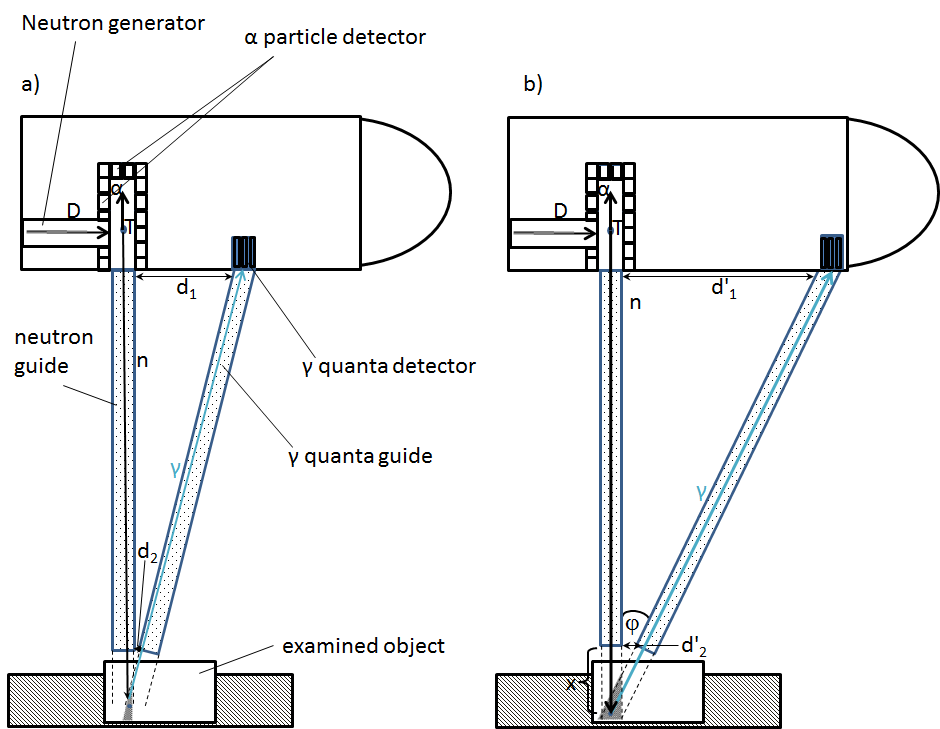}
\caption{a)~Schematic view of a device for underwater detection of hazardous substances using neutron beams.
Application of guides reduce the scattering of both neutrons and $\gamma$ quanta, while registration of an
$\alpha$ particle in anti-coincidence with gamma quanta detector allows to reduce background. b)~Demonstration
of how changing of the relative position between neutron and $\gamma$ quanta guides provides a tomographic image
of the interrogated object.}
\label{Fig1}
\end{figure}

Scheme of the proposed device is presented in Fig~\ref{Fig1}.
Neutron generator collides deuterium ions with a Tritium target producing a neutron and an $\alpha$ particle.
Because of the much higher energy released in this reaction compared to the energy of deuterium, both
particles are produced almost isotropically and move back-to-back. The $\alpha$ particle is detected
by a system of position sensitive detectors, e.g. silicon pads or scintillation hodoscope, disposed
on the walls of the generator. Neutrons emitted towards the subject of interrogation fly inside a guide
built out of a stainless steal pipe filled with low pressure air or some other gas having low cross section
for neutron interaction. Neutrons after leaving a guide may be scattered inelastically
on atomic nuclei in the tested substance. The nuclei deexcite and emit gamma quanta with
energy specific to the element.
Part of the emitted $\gamma$ quanta fly towards a dedicated detector within an analogous guide
which decrease their absorption and scattering with respect to gamma quanta flying in water.
The $\gamma$ ray detector could be again a position sensitive detector measuring the energy, time and
impact point of impinging particles.
If the diameter-to-length ratio of both guides is small the depth at which $\gamma$ quanta excite nuclei
can be determined by measuring the time $\Delta t$ elapsed from the $\alpha$ particle registration
until the time of the $\gamma$ quantum registration. Generated neutrons travel with known velocity $v_n$
the distance $L_n$ from the Tritium target to the point of interaction. Similarly the gamma emitted from
the tested object fly over a distance $L_{\gamma}$ with a speed of light $c$. Thus, $\Delta t$ can be
expressed as:
\begin{equation}
\Delta t = L_n/v_n + L_{\gamma}/c - L_{\alpha}/v_{\alpha},
\end{equation}
If we know the positions of the generator target and $\gamma$ ray detector and lengths of both guides the
distance $x$ covered by neutrons from the end of the guide to the point of interaction can be calculated as:
\begin{equation}
x  = \left(\Delta t + \frac{L_{\alpha}}{v_{\alpha}} - \frac{l_{n}}{v_{n}} -\frac{l_{\gamma}}{c} \right)
\frac{c v_n \mathrm{cos\phi}}{c \mathrm{cos}\phi + v_{n}},
\end{equation}
where $l_{n}$ and $l_{\gamma}$ are the length of guides for neutrons and $\gamma$ quanta, respectively. 
Additional information on the depth of interaction is given by changing the relative position of the guides
and the angle $\phi$ between them. As it is demonstrated in Fig.~\ref{Fig1} a) and b) changing the distance
between the guides provides registration of gamma quanta emitted from different parts of the object.
This allows one to determine the density distribution of elements building
the suspected object. In order to remove background resulting from interaction of neutrons emitted in
other directions only signals registered by the $\gamma$ quanta detector in coincidence with
signals from $\alpha$ particle detectors mounted in-line with the neutron guide are saved, while the other
coincidences are discarded.
Taking into account cross section for neutron inelastic scattering with different nuclides and the detection
efficiency of $\gamma$ quanta we can reconstruct the number of atoms of each element building the
object and compare them with the known stoichiometry of hazardous substances stored in the database.
The whole detection unit can be mounted on a small submarine steered remotely from a ship.
\section{Prototype design: simulations}
In order to optimize the dimensions and relative positions of detectors and guides we have
developed dedicated open source software package written in the C++ programming language
and based on the Monte Carlo simulation methods. Our goal is to create a fast and user-friendly tool using novel
methods of geometry definition and particle tracking based on hypergraphs~\cite{infuj}.
The simulation framework is written using the C++11 standard and Open MPI library~\cite{mpi}
supporting parallel computing and it is destined for Unix-like operating systems.
The application needs to be configured with input file defining scene description (location
and shape of all objects included in simulation, as well as substances building them), and neutron
source parameters (location, number of generated neutrons and their energies).
The parameters of neutron interaction with selected
nuclei, e.g. total cross sections, neutron and $\gamma$ quanta angular distributions and multiplicities,
are parametrized as a function of neutron energy using data from the ENDF database~\cite{endf}.
Similarly, gamma quanta energies were taken from the Evaluated Nuclear
Structure Data Files (ENSDF)~\cite{ensdf}.
\begin{figure}
\centering
\includegraphics[width=6.cm]{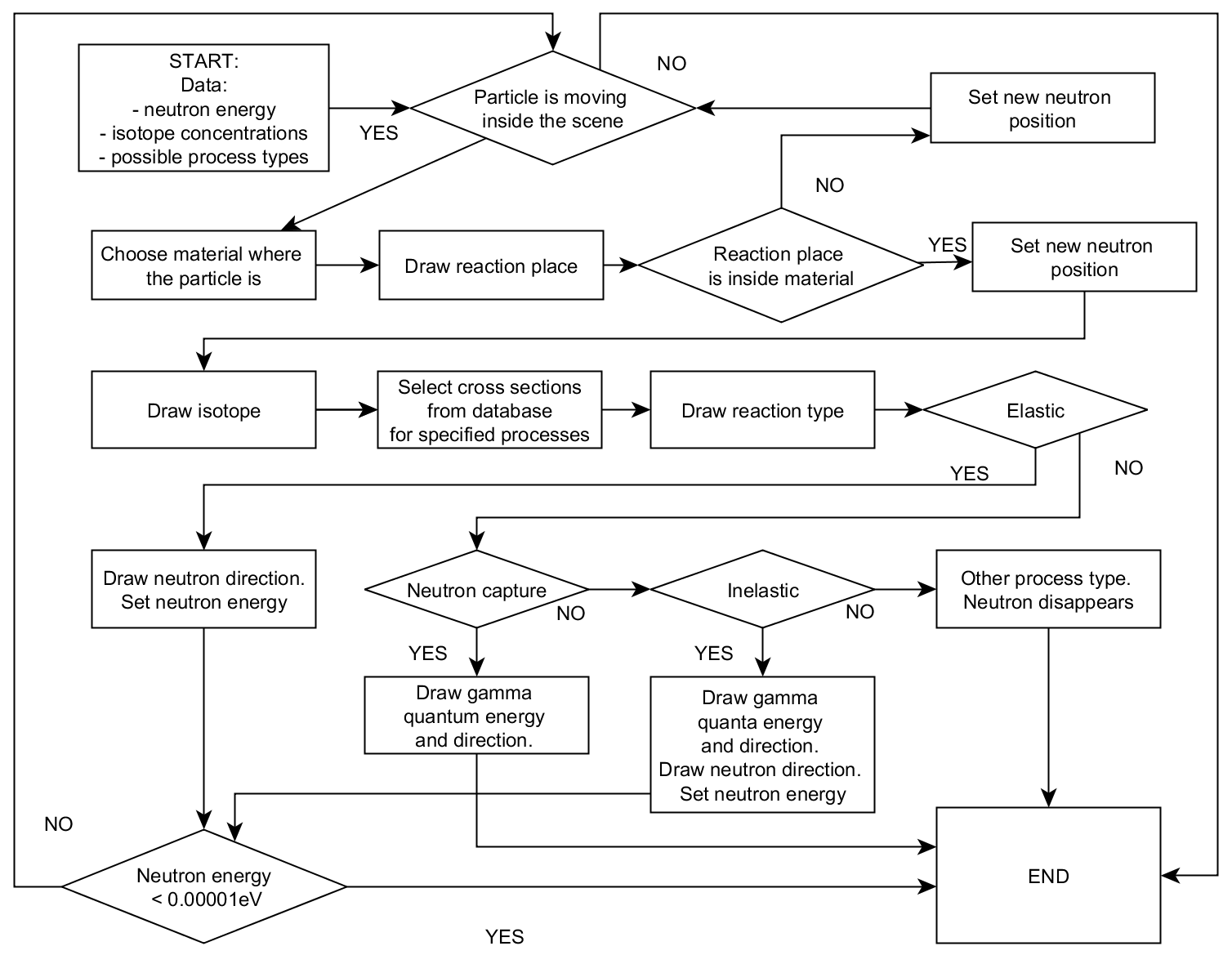}
\includegraphics[width=6.cm]{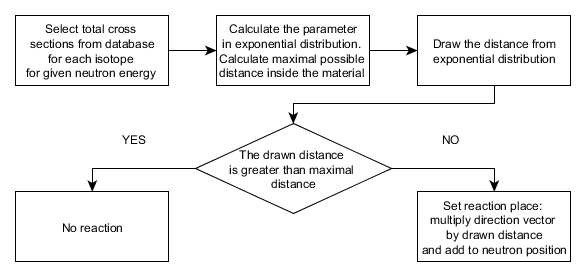}
\caption{Scheme of the algorithm used for neutron tracking (left) and details of the determination
of neutron reaction place (right).
}
\label{Fig3}
\end{figure}
This information is stored in an SQLite database and recalled during the simulation only for
elements specified by user. So far we have implemented only main processes induced by 14.1~MeV neutrons
which we are interested in, i.e. elastic and inelastic scattering and radiative capture, but in the near
future the simulations will be supplemented with other processes, e.g. fission.
At present neglected processes are taken into account effectively as one process after which
neutron is no longer tracked. Cross section for this process is calculated in a way that the
sum of all processes induced by neutron is equal to the total cross section for a given nuclei
with which the reaction occurred.
Neutrons are tracked according to the algorithm presented in Fig.~\ref{Fig3} until they
reach the scene boundary or their energy goes below the lowest included in the database~\footnote{
All results presented in this article were obtained however neglecting neutrons with energy E~$<$~10~keV.},
i.e. $10^{-5}$~eV~\cite{endf}.
The reaction place is randomly generated with exponential
probability distribution taking into account total cross section values from the database
and concentrations of all the isotopes building the substance~\footnote{
Currently we take into consideration only the most abundant isotope for 
 each element.}.
User can choose to calculate concentrations for a single chemical compound or a mixture
of substances for a specified density. Selection of an isotope with which the interaction
took place is also random and it is done again based on neutron total cross sections and
known stoichiometry of the material in which the reaction is simulated. Next the reaction
type is drawn according to cross section values for each
process from the list specified by user in the input file. The direction of neutron after
the reaction is generated using angular distributions parametrized with Legendre polynomials
or using uniform distribution if there is no relevant data in the database. It is first
determined in the center of mass system and then the four-momentum is transformed back to the
laboratory coordinate system. In case of inelastic scattering neutron energy in the laboratory
is calculated taking into account the nuclei excitation energy. In the current version of
simulations we take into account up to tenth excited nuclei level. Directions of gamma quanta
coming from the nuclei deexcitation are currently generated uniformly in the laboratory frame.
\\
As a starting point for design of the device for underwater threats detection we have defined
a simple setup with point-like source generating 14.1~MeV neutrons uniformly in space.
The scheme of the simulated setup with superimposed points of of neutrons interaction are
shown in Figs.~\ref{Fig2} and~\ref{Fig2b}. 
\begin{figure}
\centering
\includegraphics[width=5.50cm]{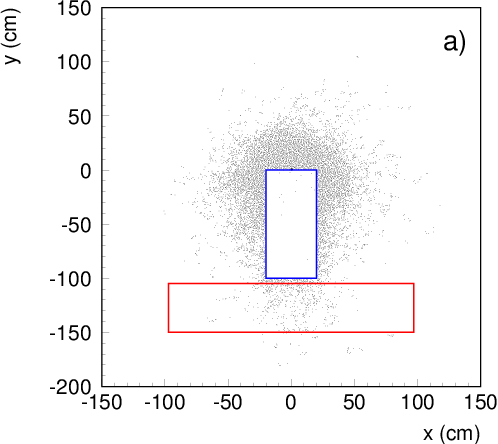}
\includegraphics[width=5.50cm]{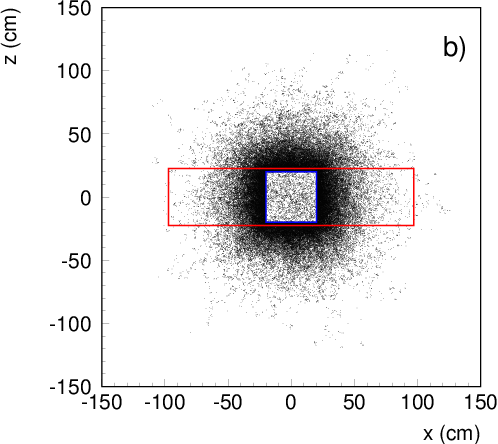}
\includegraphics[width=5.50cm]{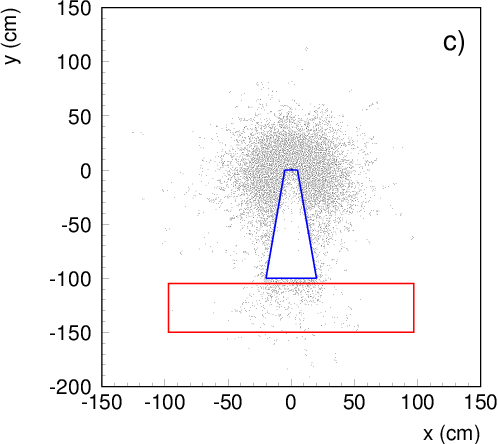}
\includegraphics[width=5.50cm]{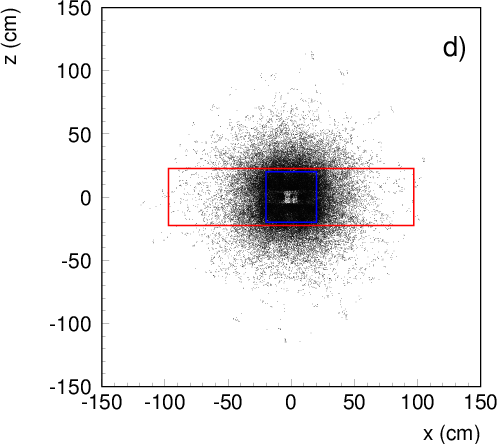}
\caption{Distribution of neutron interaction points in x-y view for guide in
a shape of a) cuboid  and b) truncated pyramid.
The point-like source of neutrons is located at the origin of the reference frame. The neutron
guide is presented as the blue rectangle (a) or trapezoid (c) aligned along the y axis.
The container with explosives is represented by the red cuboid aligned horizontally.
For better visibility of the influence of neutron guide plots a, c and b, d are made
only for -5~$\leq$~z~$\leq$~5 and y~$\leq$~0, respectively.
}
\label{Fig2}
\end{figure}
\begin{figure}
\centering
\includegraphics[width=5.50cm]{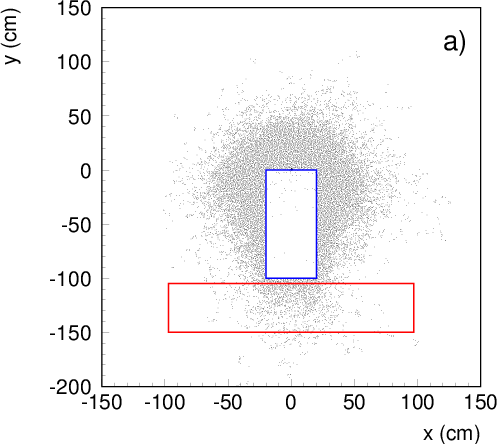}
\includegraphics[width=5.50cm]{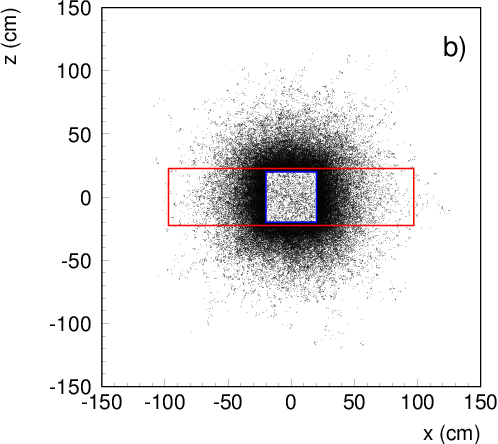}
\includegraphics[width=5.50cm]{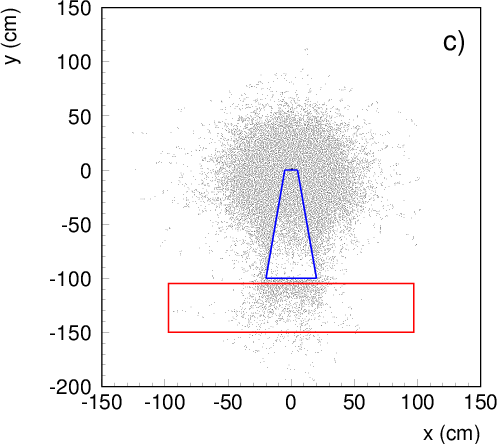}
\includegraphics[width=5.50cm]{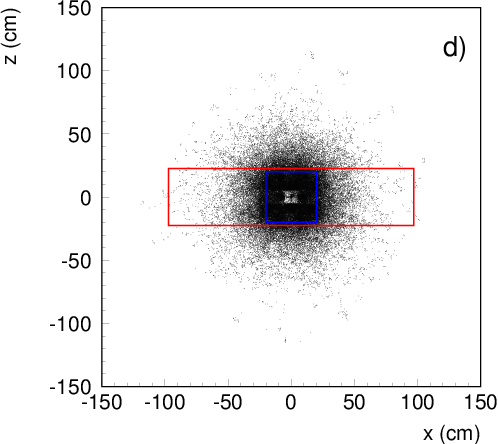}
\caption{Distribution of neutron interaction points in x-y view for guide in
a shape of a) cuboid  and b) truncated pyramid.
The point-like source of neutrons is located at the origin of the reference frame. The neutron
guide is presented as the blue rectangle (a) or trapezoid (c) aligned along the y axis.
The container with explosives is represented by the red cuboid aligned horizontally.
For better visibility of the influence of neutron guide plots a, c and b, d are made
only for -20~$\leq$~z~$\leq$~20 and y~$\leq$~0, respectively.
}
\label{Fig2b}
\end{figure}
\begin{figure}
\centering
\includegraphics[width=5.50cm]{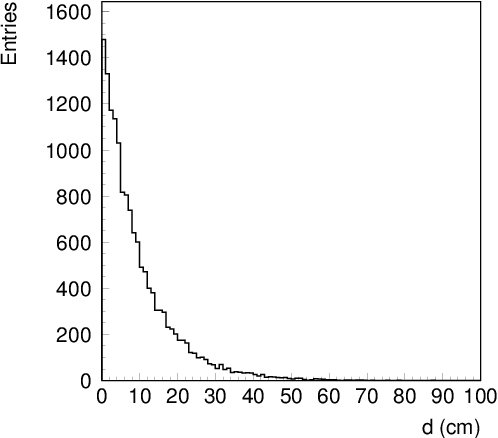}
\caption{Distribution of the distance covered by neutron till the point of first interaction in
the water (for y~$\geq$~0).
}
\label{Fig5}
\end{figure}
We have considered two neutron guides filled with air under normal conditions: cuboid with
dimensions 40~x~40~x~100~cm$^3$ (Figs.~\ref{Fig2}a and~\ref{Fig2b}a) and truncated pyramid
with height equal to 100~cm and bases with dimensions of 5~cm and 20~cm (Figs.~\ref{Fig2}b
and~\ref{Fig2b}b).
The interrogated object with dimensions 194~x~255~x~50~cm$^3$
lies on the bottom of a sea and contains mustard gas. As one can see in Fig.~\ref{Fig5}
the distribution of the path length of neutrons in water is characterized by a mean free path
of about 9.5~cm. At the same time the flux flying through the guide filled with air reaches
the container with mustard gas and excites its nuclei.
Comparing Fig.~\ref{Fig2}a and Fig.~\ref{Fig2}b one can see also that both shapes
of the neutron guide give effectively the same flux irradiating the gas container,
but for the trapezoidal shape a better spatial separation between regions where neutrons interact
in water and in the interrogated object is clearly visible. The optimization of shapes and
configuration of the neutron and gamma quanta guides will be a subject of future investigations.
\begin{figure}
\centering
\includegraphics[width=5.50cm]{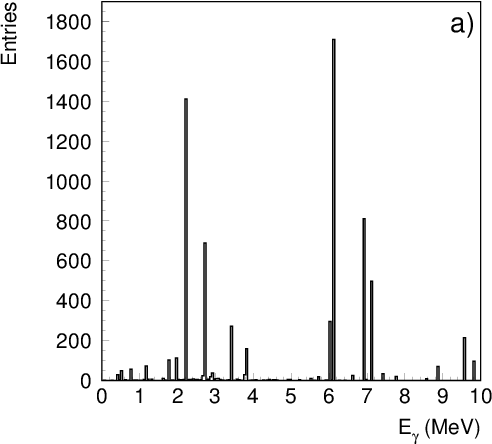}
\includegraphics[width=5.50cm]{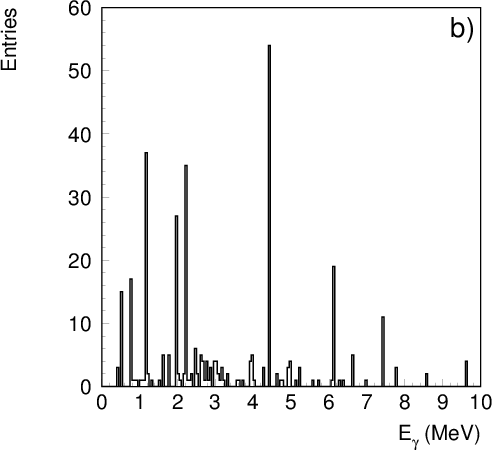}
\caption{Distributions of energies of gamma quanta coming from nuclei excitation in a) water
and b) mustard gas.
}
\label{Fig4}
\end{figure}
The energy spectra of gamma quanta from deexcitation of nuclei in water and interrogated
object are presented in Fig.~\ref{Fig4}. In Fig.~\ref{Fig4}a one can see strong lines
from the excitation of Oxygen with energies about 3~MeV, 6~MeV and 7~MeV, and characteristic
line of neutron capture by Hydrogen (about 2.2~MeV). The spectrum for the container with
mustard gas in Fig.~\ref{Fig4}b contains a big peak for gamma quanta emmited
by Carbon $^{12}$C ($E_{\gamma} =$~4.43~MeV)
and a structure at low energies and a peak at $E_{\gamma} \sim$~6~MeV comming from
Chlorine $^{37}$Cl. Hydrogen and Sulfur $^{32}$S compose one line at $E_{\gamma} =$~2.2~MeV,
which shows that identification of this element will be difficult. 
\section{Summary and outlook}
Methods of chemical thread detection based on neutron activation have a huge potential and
may open a new frontier in homeland security.
In the aquatic environment application of this method encounters serious problems since
neutrons are strongly attenuated. In order to suppress this attenuation and to decrease
background from gamma radiation induced in the water we propose to use guides for neutrons
and gamma quanta which speeds up and simplifies identification.
Moreover, it may provide a determination of the density distribution of a dangerous substance.
For designing of a device exploiting this idea we have been developing a fast
and user-friendly simulation package using novel methods of geometry definition and particle
tracking based on hypergraphs. Although we are in a very early stage of the development 
the first results indicate that indeed the guides will increase the performance of underwater
threats detection with fast neutrons.
\section*{Acknowledgments}
This work was supported by the Polish Ministry of Science and Higher Education through
grant No. 7150/E-338/M/2014 for the support of young researchers and PhD students of the
Department of Physics, Astronomy and Applied Computer Science of the Jagiellonian University.
%
%

\end{document}